\documentclass[twocolumn,floats,showpacs,prb]{revtex4}
\usepackage{amsmath}
\usepackage{graphicx}

\begin{document}

\title{A Real Space Description of the Superconducting and Pseudogap Phase}
\author{X. Q. Huang$^{1,2}$}
\email{xqhuang@netra.nju.edu.cn} \affiliation{$^1$Department of
Telecommunications Engineering ICE, PLAUST, Nanjing 210016,
China \\
$^{2}$Department of Physics and National Laboratory of Solid State
Microstructure, Nanjing University, Nanjing 210093, China }
\date{\today}

\begin{abstract}
In this work, we study the relationship between the superconducting
phase and pseudogap phase in a real-space picture. We suggest that
the superconducting ground states are guaranteed by the energy
minimum charge structure of the quasi-one-dimensional Peierls chains
(static vortex lines). It is shown that there is a charge ordering
phase transition from the Peierls chains (the superconducting ground
state) to the periodic chains (the superconducting excited state) in
any superconductors. In our scenarios, all the superconducting
electrons can be considered as the  \textquotedblleft inertial
electrons\textquotedblright\ at some stable zero-force positions.
Furthermore, we prove analytically that two electrons,  due to a
short-range real space Coulomb confinement effect (the
nearest-neighbor electromagnetic interactions), can be in pairing
inside a single plaquette with four negative ions. This implies that
the pseudogap phenomenon can be found from a wide variety of
materials, not just the cuprate superconductors.

\end{abstract}

 \maketitle

\section{Introduction}

In 1957, Bardeen, Cooper, and Schrieffer published the
 well-known microscopic theory of superconductivity.\cite{bcs} Since then most physicists
 have come to believe that the superconducting state involve electron pairs bound together by
the exchange of  phonons (atomic lattice vibrations).  In the BCS
framework, it is the lattice vibrations that provides the binding
energy of electrons in the Cooper pairs. This great theory has been
intensively challenged by the discovery of cuprate superconductors
with a critical temperature as high as 164 K,
\cite{bednorz,mkwu,gao} some theoretical condensed matter physicists
have started to doubt the reliability of the phonon-exchange-pairing
superconducting mechanism.\cite{anderson0} They consider that
phonons should be effectively ruled-out as the underlying cause of
high-temperature superconductivity in cuprates. Consequently, many
alternative quasiparticles with energies higher than the phonon
frequencies have been proposed as the reason causes the loss of
electrical resistance at the higher temperatures. In fact, these
efforts have caused much more controversy about how can strongly
repulsive electrons  form a condensate that flows without
resistance.

In the field of superconducting,  the phenomenon of
superconductivity is normally interpreted in two physical spaces:
the momentum space and the real space. In our opinion, it is more
reliable to discuss physical problems in the real space where
electron-electron and electron-ion interactions can be illustrated
in a very direct and clear manner. However, researchers seem prefer
to carry out all their study  in the momentum-space (dynamic
screening). Besides, in order to ensure the authenticity and
reliability of the physical description, there should not be any
essential difference between the real-space picture and
momentum-space picture. As we know, the BCS formalism was
established in the momentum-space where the superconducting
electrons are paired in a coherent order quantum state, while in
real-space these electrons are in a disorder phase. The fundamental
differences between two physical pictures of superconducting
electrons reveal that the momentum-space BCS theory is not the final
theory of the superconductivity. More importantly, Coulomb
interaction is the elementary electrical force that causes two
negative electrons to repel each other, furthermore, the random
interactions between electrons and lattice ions can not be ignored.
But, from the perspective of the real-space dynamics, any small
differences in force applied to each electron of a Cooper pair could
lead to the breakdown of pairing correlations. So how can the
real-space repulsion between electrons and the electron-ion
interactions be eliminated to support the formation of the Cooper
pairs? Obviously, the BCS theory of momentum-space cannot avoid this
crucial challenge.

With the invention and the application of the modern experimental
techniques, for example the Scanning Tunneling Microscopy (STM),
\cite{hanaguri} researchers now can \textquotedblleft look
inside\textquotedblright\ into the superconductors and
\textquotedblleft see\textquotedblright\ the superconducting
electrons. Motivated by these new results, more and more researchers
have tried replace the conventional superconducting picture (dynamic
screening) with a real space picture where the superconducting
electrons  congregate in some quasi-one-dimensional rivers of charge
(or vortex lines, stripes) separated by insulating domains.
\cite{hanaguri,kivelson,tranquada,komiya} Obviously, to construct a
proper model and theory related to the formation of the
one-dimensional charge rivers will be a major challenge for those
devoting themselves to crack the mystery of high-temperature
superconductivity. In recent years, we have proposed a real space
superconducting mechanism which provide new insights into the nature
of the charge stripes. \cite{huang}

In the present paper, we provide a complete theoretical analysis of
how can the Coulomb repulsion between electrons be eliminated in
favor of electron pairing and superconducting. Moreover, the
physical nature of pseudogap phenomenon and  $d$-wave pairing
symmetry are also naturally understood on the basis of the
short-range real space Coulomb confinement effect.

\section{How can the repulsion between electrons be eliminated?}

\begin{figure}[tbp]
\begin{center}
\resizebox{1\columnwidth}{!}{
\includegraphics{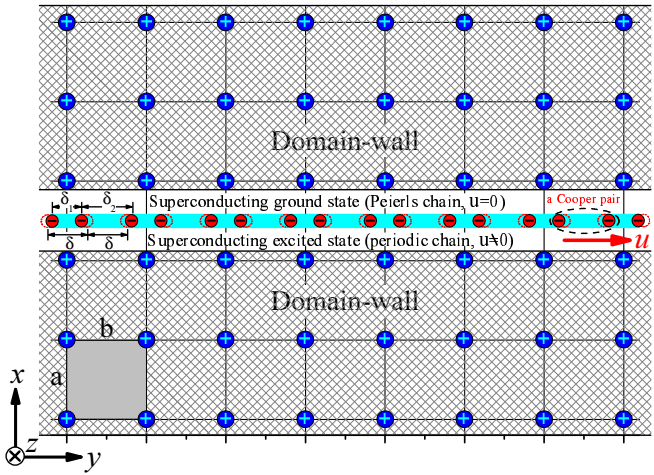}}
\end{center}
\caption{According to the energy minimum principle, the charge
carriers (electrons) will automatically gather in some zero-force
quasi-one-dimensional spaces. The Peierls chain [$\delta _{1}\neq
\delta _{2}(=b-\delta _{1}]$, where $b$ is the lattice constant)
corresponds to the superconducting ground state with
electron-transfer velocity $u=0$, while the periodic chain
corresponds to a  superconducting excited state with $u>0$. }
\label{fig1}
\end{figure}

In the momentum space (dynamic screening), the superconducting
electrons are randomly distributed in the real-space at the same
time. This implies that there doesn't exist two itinerant electrons
(the so-called pairing electrons) with the exactly the same
interaction environment. Hence, the Coulomb interaction between two
electrons (Cooper pair) of the dynamic screening can not be
completely suppressed by the other electrons and ions. Now, how can
the repulsion between electrons be suppressed to support the
occurrence of superconductivity? The answer lies in the
quasi-one-dimensional real-space stripes, as shown in Fig.
\ref{fig1}.

In the previous paper, \cite{huang} we proved theoretically that a
static one-dimensional vortex line (Peierls chain, or charge stripe)
can be naturally formed inside the superconducting plane. The stable
charge stripe is confined by the domain-walls of the positive ions
(see Fig. \ref{fig1}). Driven by an external electric field, there
will occur a charge ordering phase transition from the stable
Peierls chain (the superconducting ground state of $u=0$) to the
periodic chain (the superconducting excited state of $u>0$) in any
superconductors, as illustrated in Fig. \ref{fig1}. For
three-dimensional superconductors, the vortex lines can further
self-organize into some thermodynamically stable vortex lattices
with trigonal or tetragonal symmetry.

\begin{figure}[tbp]
\begin{center}
\resizebox{1\columnwidth}{!}{
\includegraphics{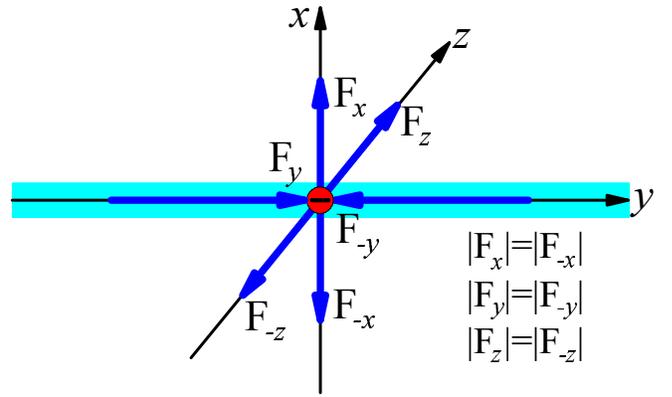}}
\end{center}
\caption{The external electromagnetic forces be exerted on one of
the electrons in Fig. \ref{fig1}. In our scenarios, all the
superconducting electrons should be in exactly the same inertial
state. In this case, the forces on any one of them should be
balanced as shown in this figure.} \label{fig2}
\end{figure}

In any superconducting materials, to a very good approximation,
there are mainly two kinds of Coulomb interactions on
superconducting electron: (1) electron-electron repulsive
interactions, and (2)electron-ion attractive interactions. In our
superconducting scenarios, to maintain the zero resistance property
of the superconductors it is necessary that all superconducting
electrons should be in the zero-force state, or electron is the
\textquotedblleft inertial electrons\textquotedblright. Figure
\ref{fig2} shows the forces applied to one of the electrons in Fig.
\ref{fig1}. Due to the crystal symmetry in $x$ and $z$ directions,
all the forces (electron-ion interactions) in these two directions
are canceled in opposite directions. This can be expressed
mathematically as follows:
\begin{equation}
\left\vert \mathbf{F}_{x}\right\vert =\left\vert
\mathbf{F}_{-x}\right\vert
,\quad \left\vert \mathbf{F}_{z}\right\vert =\left\vert \mathbf{F}%
_{-z}\right\vert .  \label{fxz}
\end{equation}
In the $y$-direction of charge stripe, the forces on the electron
come from the electron-electron interactions inside the charge
stripe. In our scenarios, when a superconductor enters into the
superconducting state, all the superconducting electrons can be
considered as the \textquotedblleft inertial
electrons\textquotedblright\ and forces on any one of them should be
balanced. Hence we have
\begin{equation}
\left\vert \mathbf{F}_{y}\right\vert =\left\vert
\mathbf{F}_{-y}\right\vert .  \label{fxz}
\end{equation}

\begin{figure}[tbp]
\begin{center}
\resizebox{1\columnwidth}{!}{
\includegraphics{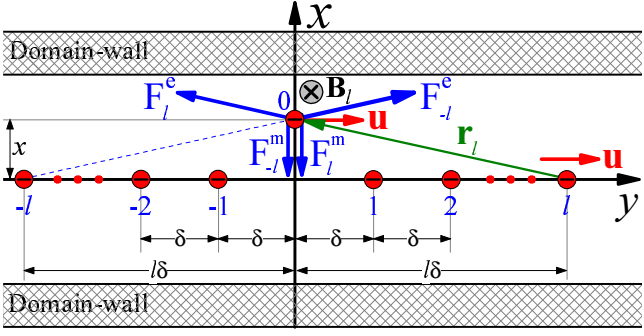}}
\end{center}
\caption{The electron \textquotedblleft 0\textquotedblright\ with a
small displacement in $x$-direction ($x \ll \delta$) will experience
a repulsive electrostatic force and an attractive magnetic force by
the other electrons in the charge stripe.  } \label{fig3}
\end{figure}

According to the symmetry of the charge stripe, it is not difficult
to find that the condition above can be satisfied naturally without
depending on any \textquotedblleft glue\textquotedblright.
Generally, for an infinite one-dimensional superconducting periodic
chain with the electron-electron spacing $\delta=b/2$ and
electron-transfer velocity $u>0$ (as shown in Fig. \ref{fig1}). Now,
let us suppose one of the electrons (marked by \textquotedblleft
0\textquotedblright\ in Fig. \ref{fig3}) has a small displacement in
$x$-direction ($x \ll \delta$), if the velocity $u=0$, the other
electrons inside the stripe will repel the electron
\textquotedblleft 0\textquotedblright\ through the electrostatic
field.  The repulsive forces from the symmetry electron pair
$(l,-l)$ of Fig. \ref{fig3} can be expressed as:
\begin{eqnarray}
\mathbf{F}^{e}(l,-l) &=&-e(\mathbf{E}_{l}^{e}+\mathbf{E}_{-l}^{e})
\nonumber
\\
&=&\frac{e^{2}\cos \theta }{2\pi \varepsilon _{0}r_{l}^{2}}\mathbf{i=}\frac{%
e^{2}x}{2\pi \varepsilon _{0}r_{l}^{3}}\mathbf{i}, \label{force_e}
\end{eqnarray}%
where $\mathbf{E}_{l}^{e}$ and $\mathbf{E}_{-l}^{e}$ are the
electrostatic field generated by the electron $l$ and electron $-l$,
respectively, and $\mathbf{i}$ is the unit vector in $x$-direction.

When the velocity $u\neq 0$, apart from the repulsive electrostatic
force,  the electron $0$ will experience an attractive magnetic
force exerted by the other electrons. According to electromagnetic
theory, we can find that the magnetic field generated by the
electron $l$ around the by the electron $0$ can be expressed as:
\begin{equation}
\mathbf{B}_{l}=-\frac{\mu _{0}e\mathbf{u}\times \mathbf{r}_{l}}{4\pi
r_{l}^{3}}.  \label{magnetic}
\end{equation}%

The above magnetic field will exert on the moving electron $0$ with
the following Lorentz force:
\begin{eqnarray}
\mathbf{F}_{l}^{m} &=&-e\mathbf{u}\times \mathbf{B}_{l}  \nonumber \\
&=&-\frac{\mu _{0}e^{2}u^{2}x}{4\pi r_{l}^{3}}\mathbf{i=}-\frac{\mu
_{0}e^{2}u^{2}x}{4\pi (x^{2}+l^{2}\delta ^{2})^{3/2}}\mathbf{i.}
\label{force_m}
\end{eqnarray}

Similarly, we can define a pair of magnetic force on the electron
$0$ as follow
\begin{equation}
\mathbf{F}^{m}(l,-l)=\mathbf{F}_{l}^{m}+\mathbf{F}_{-l}^{m}=2\mathbf{F}%
_{l}^{m}=-\frac{\mu _{0}e^{2}u^{2}x}{2\pi r_{l}^{3}}\mathbf{i}.
\label{force_m1}
\end{equation}%

From Eq. (\ref{force_e}) and Eq. (\ref{force_m1}), we have the
following representation:
\begin{equation}
\left\vert \frac{\mathbf{F}^{e}(l,-l)}{\mathbf{F}^{m}(l,-l)}\right\vert =%
\frac{c^{2}}{u^{2}}.  \label{force_em}
\end{equation}%

Because $u<c$, thus from  Eq. (\ref{force_em}) it follows that the
magnetic attraction $\mathbf{F}^{m}(l,-l)$ is always less than the
electric repulsion $\mathbf{F}^{e}(l,-l)$.

When $\delta (=b/2)\gg x$, then we have $r_{l}\approx l\delta $.
Now, combining  Eq. (\ref{force_e}) and Eq. (\ref{force_m1}), we
have the total electromagnetic force on the electron:
\begin{eqnarray}
F_{Total} &=&\sum_{l=1}^{\infty }\left\vert \mathbf{F}^{e}(l,-l)+\mathbf{F}%
^{m}(l,-l)\right\vert   \nonumber \\
&=&\frac{e^{2}x}{4\pi \varepsilon _{0}\delta ^{3}}\left( 1-\frac{u^{2}}{c^{2}%
}\right) \sum_{l=1}^{\infty }\frac{1}{l^{3}}  \nonumber \\
&=&\frac{2e^{2}\varsigma (3)}{\pi \varepsilon _{0}b^{3}}\left( 1-\frac{u^{2}%
}{c^{2}}\right) x.  \label{forcet}
\end{eqnarray}
where $\varsigma (3)$ is a zeta function.

It can be easily found from Eq. (\ref{forcet}),  if the displacement
$x \neq 0$, the $F_{Total}$ is always positive. Then from the view
point of energy, this repulsive force will lead directly to an
increase in the system energy, which in turn decrease the stability
of the superconducting state. But, it is clear that if $x=0$, we
have $F_{Total}=0$ according to Eq. (\ref{forcet}). This result
indicates that a perfect one-dimensional charge-chain can not only
reduce the energy of the superconducting state, but also eliminate
completely the repulsion between the electrons inside the charge
stripe.

\begin{figure}[bp]
\begin{center}
\resizebox{1\columnwidth}{!}{
\includegraphics{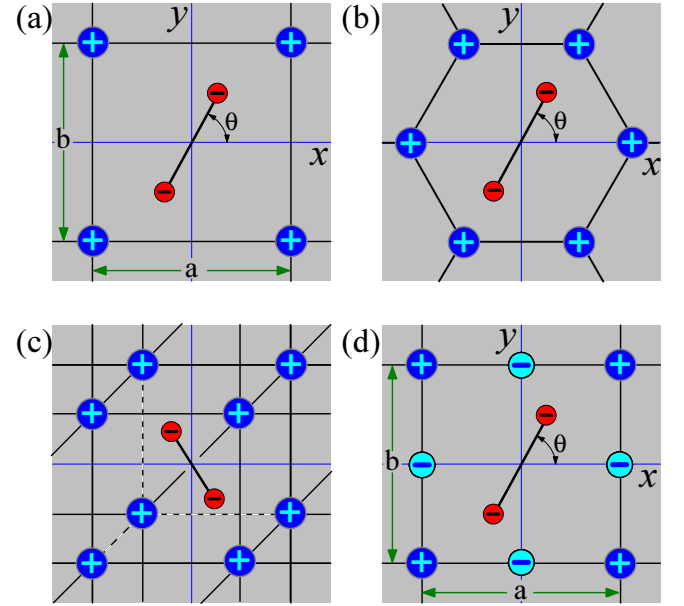}}
\end{center}
\caption{ Two electrons (a Cooper pair) inside a unit cell of
different superconductors. (a) Two-dimensional rectangular or square
, (b) two-dimensional hexagonal structure (for example, $MgB_{2}$),
(c) three dimensional simple cubic crystal (or body-centered cubic,
face-centered cubic), (d) two-dimensional Cu-O plane. We prove that
the Cooper pair can survive only in (d) with the nearest-neighbor
negative ions.} \label{fig4}
\end{figure}

\section{ Pseudogap State}

One of the highly controversial issues in the understanding of
high-temperature superconductivity is the origin of the normal-state
gap (pseudogap). There are many theories and models attempt to
describe the pseudogap phase, yet its nature remains a mystery. Here
we present a new interpretation of pseudogap based on the simple
real-space Coulomb confinement effect.

In the framework of the real-space approach (as discussed in the
above section), the most important unit of the superconducting phase
is the one-dimensional long-range correlated charge stripes that may
form through long-range lattice confinement effect
(electron-electron and electron-ion interactions). However, we will
show that the nearest-neighbor electron-ion correlation may be
responsible for the mechanism of the pseudogap. Figure \ref{fig4}
shows two electrons (a Cooper pair) inside a unit cell of different
superconductors.  For the cases of Figs. \ref{fig4}(a)-(c), it is
not difficult to find that the  \textquotedblleft Cooper
pair\textquotedblright\ will be split up due to electron-ion
interactions, while the Cooper pair can survive in Fig.
\ref{fig4}(d) in two special directions.

In what follows, we pay our attention to the case of Fig.
\ref{fig4}(d). We try to show how can two repulsive electrons stay
together inside a single plaquette and discuss the pairing symmetry
of the corresponding pseudogap phase. Here, we consider only two
specific situations where two electrons ($A$ and $B$ of Fig.
\ref{fig5}) arranged on a line in $y$ and $xy$-direction, as shown
in Figs. \ref{fig5}(a) and (b) respectively. As can be seen from the
figure, there are four nearest-neighbor ions (marked by 1, 2, 3, 4)
with a negative charge ($Q^{-}=-me$) and four next-nearest-neighbor
ions (marked by 5, 6, 7, 8) with a positive charge ($Q^{+}=ne$)
around the electron pair. In these two cases, because of the
structural symmetry, we can present the explicit analytical
expressions of the confinement forces on the electrons. Figures \ref%
{fig5}(a) and (b) illustrate the eight Coulomb forces ($f_{1}$,
$f_{2}$, $f_{3}$, $f_{4}$, $f_{5}$, $f_{6}$, $f_{7}$ and $f_{8}$)
exerted on the electron $A$ by the ions and the repulsive force
($f_{B}$) inside the pair. Based on Fig. \ref{fig5}(a) (For the sake
of simplicity, suppose $a=b$), we can get a general formula of the
total confinement force $F_{y}$ applied to the electron $A$ in
$y$-direction as:
\begin{figure}[tbp]
\begin{center}
\resizebox{1\columnwidth}{!}{
\includegraphics{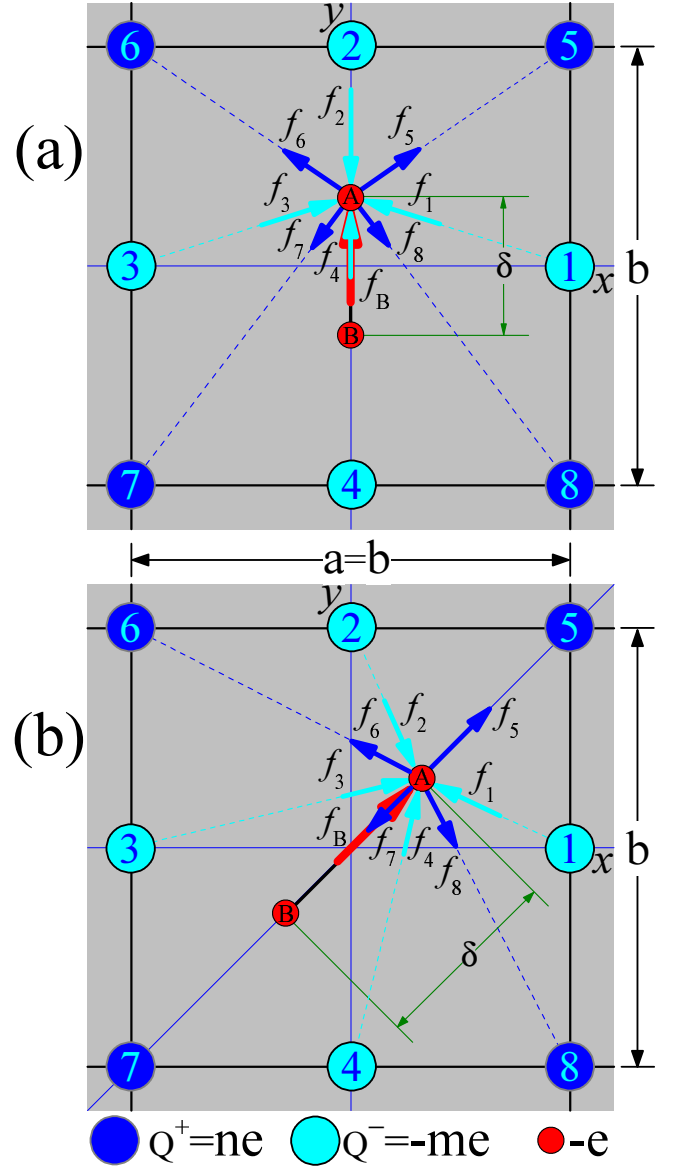}}
\end{center}
\caption{The schematic plot of the confinement forces acting the
electron pair ($A$ and $B$) inside one unit cell of the
superconducting plane. For each electron, there are four negative
nearest-neighbor ions (marked by 1, 2, 3, 4) with a charge
($Q^{-}=-me$) and four next-nearest-neighbor ions (marked by 5, 6,
7, 8) with ($Q^{+}=ne$) which will exert forces on it. Two special
situations are considered in this study, they are (a) two electrons
(a Cooper pair) arranged along the $y$-direction, and (b) along the
$xy$-direction. } \label{fig5}
\end{figure}
\begin{equation}
F_{y}=f_{B}+F_{y}^{(1)}+F_{y}^{(2)}.  \label{force_y}
\end{equation}%
The well-known Coulomb repulsion $f_{B}$ can be represented as
\begin{equation}
f_{B}=\frac{e^{2}}{4\pi \varepsilon _{0}\delta ^{2}}, \label{kl}
\end{equation}%
where $\delta$ is the electron-electron spacing which can be used to
characterize the size of the Cooper pair.

$F_{y}^{(1)}$ is the total nearest-neighbor Coulomb force which is
defined as
\begin{eqnarray}
F_{y}^{(1)} &=&(f_{1}+f_{3})-(f_{2}+f_{4})  \nonumber \\
&=&m\frac{e^{2}}{4\pi \varepsilon _{0}}\left( \frac{1}{d_{1}^{2}}-\frac{1}{%
d_{2}^{2}}\right) ,  \label{nnb}
\end{eqnarray}%
and the total next-nearest-neighbor Coulomb force $F_{y}^{(2)}$ can
be expressed as
\begin{eqnarray}
F_{y}^{(2)} &=&(f_{5}+f_{6})-(f_{7}+f_{8})  \nonumber \\
&=&n\frac{e^{2}}{4\pi \varepsilon _{0}}\left( \frac{1}{d_{3}^{2}}-\frac{1}{%
d_{4}^{2}}\right) .  \label{snnb}
\end{eqnarray}%%
The parameters $d_{1}$, $d_{2},$ $d_{3}$ and $d_{4}$ are given by
\begin{eqnarray}
\quad d_{1} &=&\frac{\left( b^{2}+\delta ^{2}\right) ^{3/4}}{2\sqrt{2\delta }%
},\quad d_{2}=\frac{b^{2}-\delta ^{2}}{4\sqrt{b\delta }}, \nonumber
\\
d_{3} &=&\frac{(2b^{2}+\delta ^{2}-2b\delta )^{1/4}\sqrt{2b^{2}+\delta ^{2}-%
\sqrt{2}b\delta }}{4\sqrt{b-\delta }},  \nonumber \\
d_{4} &=&\frac{(2b^{2}+\delta ^{2}+2b\delta )^{1/4}\sqrt{2b^{2}+\delta ^{2}+%
\sqrt{2}b\delta }}{4\sqrt{b+\delta }}.  \label{d_14}
\end{eqnarray}%

Similarly, in $xy$-direction, we have
\begin{equation}
F_{xy}=f_{B}+F_{xy}^{(1)}+F_{xy}^{(2)}, \label{force_xy}
\end{equation}%
and the corresponding functions are defined by
\begin{eqnarray}
f_{B} &=&\frac{e^{2}}{4\pi \varepsilon _{0}\delta ^{2}},  \label{fb2} \\
F_{xy}^{(1)} &=&m\frac{e^{2}}{4\pi \varepsilon _{0}}\left( \frac{1}{D_{1}^{2}%
}-\frac{1}{D_{2}^{2}}\right) , \quad  m=1, 2, 3...  \label{f_xy1} \\
F_{xy}^{(2)} &=&n\frac{e^{2}}{4\pi \varepsilon _{0}}\left( \frac{1}{D_{3}^{2}%
}-\frac{1}{D_{4}^{2}}\right), \quad  n=1, 2, 3...  \label{f_xy2}
\end{eqnarray}%
The four distance parameters above are given by
\begin{eqnarray}
D_{1} &=&\frac{(b^{2}+\delta ^{2}+\sqrt{2}b\delta )^{3/4}}{2\sqrt{\sqrt{2}%
b+2\delta }},\quad   \nonumber \\
D_{2} &=&\frac{(b^{2}+\delta ^{2}-\sqrt{2}b\delta )^{3/4}}{2\sqrt{\sqrt{2}%
b-2\delta }},  \nonumber \\
D_{3} &=&\frac{(2b^{2}-\delta ^{2})}{4\times 2^{3/4}\sqrt{b\delta
}},\quad D_{4}=\frac{\left( 2b^{2}+\delta ^{2}\right)
^{3/4}}{4\sqrt{\delta }},\quad \label{D1_4}
\end{eqnarray}%
where $\delta <a/\sqrt{2}.$

\begin{figure}[tbp]
\begin{center}
\resizebox{1\columnwidth}{!}{
\includegraphics{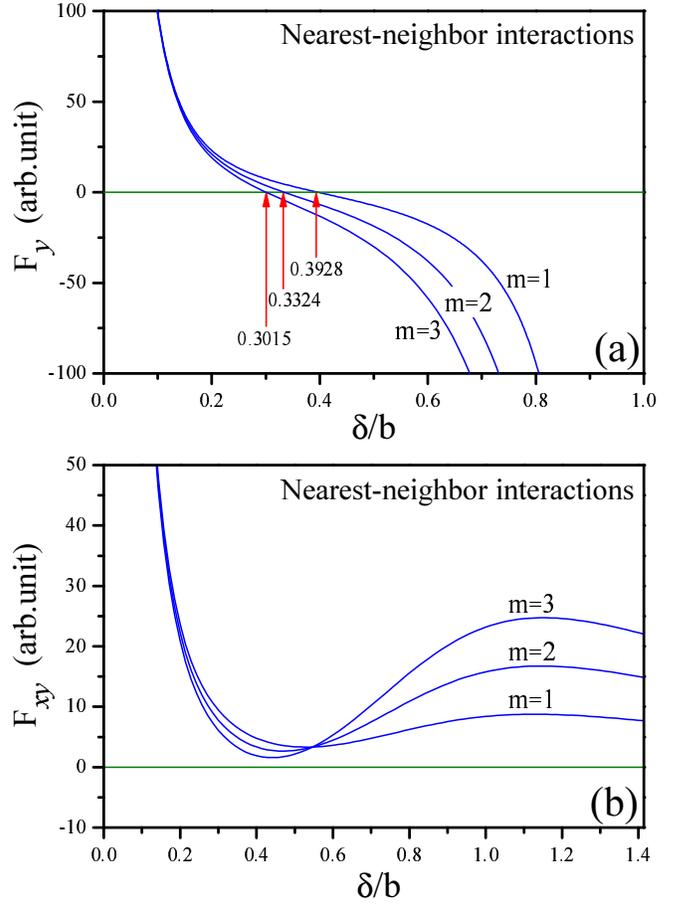}}
\end{center}
\caption{Analytical confinement forces versus $\delta /b$ in two
special directions. (a) Two electrons (a Cooper pair) are arranged
in $y$-direction, in this case, a stable real-space Cooper pair can
be formed inside the unit cell due to the existence of $F_{y}=0$,
(b) while two electrons in $xy$-direction, the confinement force
$F_{xy}=0$ is always positive, indicating an unstable state of the
Cooper pair in this direction. These results implies a possibility
pseudogap phase of $d$-wave symmetry in the superconductor.}
\label{fig6}
\end{figure}

\begin{figure}[tbp]
\begin{center}
\resizebox{1\columnwidth}{!}{
\includegraphics{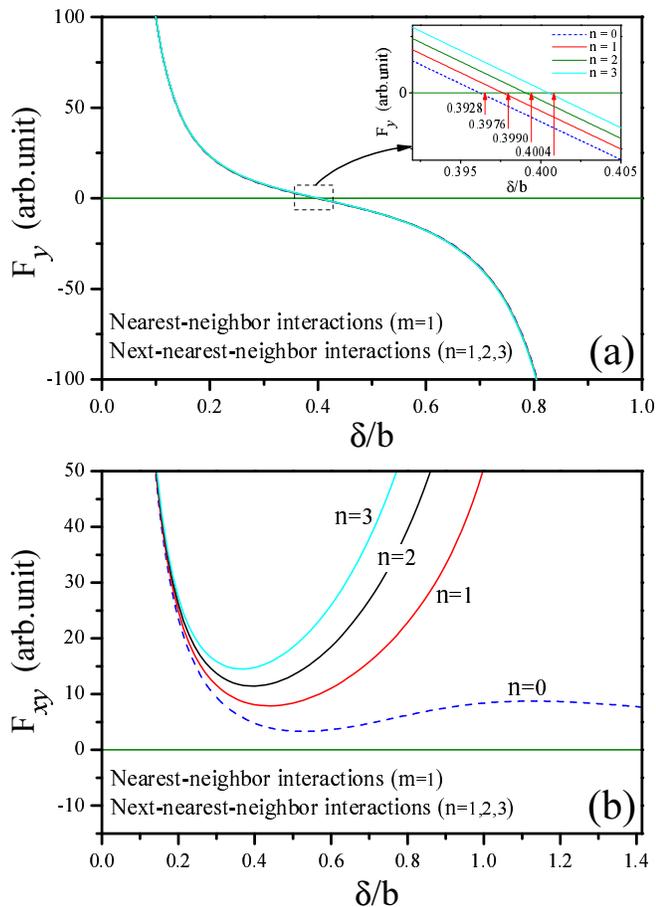}}
\end{center}
\caption{The influence of the next-nearest-neighbor interactions on
the formation of the pseudogap. (a) The adding of the
next-nearest-neighbor interactions have little impact on the
formation of electron pair in $y$-direction, the pairing electrons
can still maintain its integrity inside the single plaquette, (b)
while in the diagonal $xy$-direction, these interactions make the
pairing more difficult.} \label{fig7}
\end{figure}

In the framework of our theory, whether the two electrons become
paired (Characterized by the pseudogap) in Figs. \ref{fig5}(a) and
(b) can be judged by the value of $F_{y}$ and $F_{xy}$,
respectively. For a given superconductor with the definite structure
parameters $b$, $Q^{+}$($=ne$) and $Q^{-}$($=-me$), if there exist a
value of $\delta$ (electron-electron spacing) which can ensure
$F_{y}=0$ (or $F_{xy}=0$), then the pair can maintain it integrity
in the square lattice of Fig. \ref{fig5}(a) [or Fig. \ref{fig5}(b)].
With the analytical expressions from (\ref{force_y}) to (\ref
{D1_4}), we draw in Fig. \ref{fig6} and Fig. \ref{fig7} the total
forces ($F_{y}$ and $F_{xy}$) on the electron $A$ (a similar
discussion may be expected to be valid for the electron $B$) versus
$\delta /b$ for the cases of the nearest-neighbor and
next-nearest-neighbor interactions, respectively.

Figure \ref{fig6}(a) shows the relation between $F_{y}$ and
$\Delta/b$ under the nearest-neighbor condition. As can be seen from
the figure, there exist always one $\delta$ with the force
$F_{y}=0$. Moreover, with the increasing of the charge of the ions
$Q^{-}$ from $-e$, $-2e$ to $-3e$, the size of Cooper pair $\delta$
will decrease eventually from $0.3928b$, $0.3324b$ to $0.3015b$,
indicating a stronger confinement effect and a higher Cooper-pair
binding energy. When the two electrons arranged in $xy$-direction,
the forces $F_{xy}$ are always positive for any given $m$ (or
$Q^{-}$), as shown in Fig. \ref{fig6}(b). This result implies that
 the pair along $xy$-direction can be easily destroyed due to the
Coulomb interaction between the pair and ions. If we use the
parameter $\delta$ as a measure of the binding energy ($E_{b}\propto
1/\delta $) of the Cooper
pair, it is not difficult to conclude that the pair parallel to $x$- and $y$%
-axis has a minimum $\delta$ that leads to a maximum binding energy
in these four directions, while the binding energy may be zero when
the pair in the four diagonal directions. Hence, the
nearest-neighbor electron-ion interactions can directly lead to the
$d$-wave  pseudogap observed in the cuprate superconductors.
 To make this argument more convincing, we will take into account the
 next-nearest-neighbor interactions.

When both the nearest-neighbor and next-nearest-neighbor
interactions are present in the calculation of the force $F_{y}$ and
$F_{xy}$, we find that the real-space electron pairing symmetry
(Fig. \ref{fig7}) does not change comparing to Fig. \ref{fig6} of
the nearest-neighbor interactions. Figure \ref{fig7}(a) shows that
the adding of the next-nearest-neighbor interactions almost does not
affect the electron pair in the $y$-direction, very little
difference in the size of Cooper pair $\delta$ can be observed (see
the insert figure). In the case of Fig. \ref{fig7}(b), although
significant changes were observed in the figure, but the force
$F_{xy}$ can be guaranteed to be positive when considering different
next-nearest-neighbor interactions ($n=1,2,3$). Furthermore, the
larger $F_{xy}$ indicates that the two electrons are much more
difficult to be paired in the diagonal directions. These results
confirm our argument that the nearest-neighbor electron-ion
interactions play an important role in the origin of the pseudogap
phenomenon.

From our study of the mechanism of the pseudogap based on the
short-range real-space Coulomb confinement effect, the pair-pair
interactions have been completely neglected. This approximation is
reasonable to describe the underdoped cuprate superconductors with a
small carrier concentration. Whereas for the overdoped
superconductors, the pseudogap phase will be destroyed by pair-pair
interactions, as shown by the experimental results. Besides, the
characteristic of local nearest-neighbor interactions of our theory
implies that the pseudogap phenomenon can be found from a wide
variety of materials, especially those with low carrier
concentrations, such as the semiconductors and some insulation
materials. \cite{ stewart}

\section{Concluding remarks}

In conclusion, we have shown that the Coulomb repulsion between
electrons can be completely eliminated only if the charge carriers
(electrons) self-organize into some quasi-one-dimensional charge
stripes (vortex lines). It has been argued that the superconducting
ground states are guaranteed by the energy minimum
quasi-one-dimensional Peierls chains which are formed with the
lattice confinement effect. In the framework of our theory, all the
superconducting electrons can be considered as the \textquotedblleft
inertial electrons\textquotedblright\ at some stable zero-force
positions. Furthermore, we have proved analytically that two
electrons, due to the nearest-neighbor electromagnetic interactions,
can be in pairing inside a single plaquette with the $d$-wave
pairing symmetry. Our results suggest that the pseudogap is a common
feature that can be found from a wide variety of materials, not just
the cuprate superconductors.

\end{document}